\documentstyle[epsfig]{elsart}
\begin{document}
\runauthor{Ishida}
\begin{frontmatter}
\title{Existence of $\sigma (600)/\kappa (900)$ and 
Possible Classification of Chiral Scalar Nonet}

\author[TIT]{Muneyuki Ishida}
\address[TIT]{Department of Physics, Tokyo Institute of Technology,
Tokyo 152-8551, Japan}

\begin{abstract}
Recently, by reanalyzing the phase shift data of $I=0\ \pi\pi$ 
and of $I=1/2\ K\pi$ scatterings, we showed the evidences for
existence of light scalar mesons, $\sigma (600)$ and 
$\kappa (900)$, respectively, which had been sought but missing for a long
time. The $\sigma (600)$ and $\kappa (900)$ together with the established
resonances, $a_0(980)$ and $f_0(980)$, are shown to be consistently
classified as the members of a single scalar $\sigma$-nonet, appearing 
in the $SU(3)$ linear $\sigma$-model. Especially 
the mass value of the iso-singlet flavor-octet member
satisfies, together with those of $\kappa (900)$ and $a_0(980)$,
the Gell-Mann Okubo mass formula. 
The repulsive background phase shift $\delta_{BG}$,
which was essential to lead the $\sigma /\kappa$-existence
in our phase shift analyses,  is also shown to be 
quantitatively describable
in the framework of linear $\sigma$ model. 
Thus, the origin of $\delta_{BG}$
is reduced to the 
``compensating $\lambda\phi^4$-interaction," 
necessary from the viewpoint of chiral symmetry.
\end{abstract}
\end{frontmatter}

\section{$\pi\pi /K\pi$ phase shift analyses 
and existence of $\sigma (600)/\kappa (900)$}
We have recently made reanalyses of $\pi\pi$ and $K\pi$
phase shifts and found strong evidences for existence of $I=0$
$\sigma$ meson and $I=1/2$ $\kappa$ meson, respectively, which had been 
sought but missing for a long time. For detailed analyses, see
ref. \cite{ref1}. Through the analyses the masses and widths of 
$\sigma$ and $\kappa$, and their pole positions are determined 
with the values (in MeV),
\begin{eqnarray}  
m_\sigma &=& 585\pm 20(535\sim 675),
\Gamma_\sigma =385\pm 70,
\sqrt{s_{\rm pole}}=602\pm 26-i(196\pm 27),\ \ \ \ \ \nonumber\\
m_\kappa &=& 905\stackrel{+65}{\scriptstyle -30},
\Gamma_\kappa =545\stackrel{+235}{\scriptstyle -110},
\sqrt{s_{\rm pole}}=875\pm 75-i(335\pm 110).\ \ \ \ \ 
\end{eqnarray}
\section{Gellmann-Okubo mass formula for scalar $\sigma$-nonet}\footnote{
The content of \S 2 and \S 3 of this talk is a recapitulation
of my recent work.\cite{ref2} See also the related work.\cite{ref3}
}
Now we have scalars below 1 GeV, $\sigma (600)$ and $f_0(980)$
with $I=0$, $\kappa (900)$ with $I=1/2$ and $a_0(980)$ with $I=1$.
Among these scalar mesons,
experimentally,
$f_0(980)$ has a considerably small $\pi\pi$-width 
regardless of its large phase volume, while having a rather  
large $K\bar K$-width in spite of the fact that 
its mass is quite close
to the $K\bar K$-threshold.
Assuming approximate validity of the OZI rule, 
this fact seems to suggest that 
$f_0(980)$ consists of almost 
pure $s\bar s$-component.
Here we simply assume that $\sigma$(600) and $f_0(980)$ are the 
ideal mixing\cite{rf:ideal} states
of a single scalar nonet and that the squared-mass matrix takes 
a diagonal form in these ideal bases.
The ideal-mixing states are related 
to the octet state $\sigma_8$ and the singlet state $\sigma_1$
through the orthogonal transformation:
\begin{eqnarray}
\left(
\begin{array}{c}
\sigma (600)\\
f_0(980)
\end{array}
\right)
=
\left(
\begin{array}{c}
\sigma_n\\
\sigma_s
\end{array}
\right)
 &=& 
O
\left(
\begin{array}{c}
\sigma_8\\
\sigma_1
\end{array}
\right) ,\ \ \ \ 
\sigma_n \equiv\frac{u\bar u+d\bar d}{\sqrt{2}},\ \ 
\sigma_s \equiv s\bar s,
\label{eq:or}
\end{eqnarray}
where $O$ is the matrix of the orthogonal transformation, 
whose elements are given by
$O_{11}=O_{22}=\sqrt{1/3}$ and  $O_{12}=-O_{21}=\sqrt{2/3}$.
Through the transformation $O$,
the elements of the squared-mass matrix in the 
octet-singlet bases are numerically given by
\begin{eqnarray}
\left(
\begin{array}{cc}
 m_{\sigma_8}^2 & m_{\sigma_{81}}^2 \\
 m_{\sigma_{81}}^2 & m_{\sigma_1}^2
\end{array}
\right)  
 = 
{}^tO
\left(
\begin{array}{cc}
 m_{\sigma (600)}^2 & 0 \\
 0 & m_{f_0(980)}^2
\end{array}
\right)
O 
 =
\left(
\begin{array}{cc}
 (0.87)^2 & -(0.54)^2 \\
 -(0.54)^2 & (0.74)^2
\end{array}
\right) ,\ \ \ \ \ \ 
\label{eq:mm}
\end{eqnarray}
where the numerical values are given in (GeV)$^2$, and 
we have used the experimental values $m_\sigma$=0.59 GeV 
and $m_{f_0(980)}$=0.98 GeV.\footnote{
The fact that $m_{\sigma_1}$ is smaller than $m_{\sigma_8}$ 
is in contrast with the case 
of pseudoscalar $\eta$-$\eta '$ mass splitting. 
This fact reflects the property of the 
$U_A(1)$-breaking interaction (see, \S 3).
In this connection, note that the famous nonet mass 
formula\cite{rf:ideal} of the vector mesons,
$m_\rho^2=m_\omega^2$ and $m_\phi^2-m_{K^*}^2=m_{K^*}^2-m_\rho^2$,
is valid in the case $m_{V_1}^2=m_{V_8}^2$.
}
The mass $m_{\sigma_8}$ can also be determined theoretically by using 
the Gell-Mann Okubo (GMO) relation,
\begin{eqnarray}
m_\kappa^2 = (3m_{\sigma_8}^{\rm theor\ 2}+m_{a_0}^2)/4, 
\ \ \ 
{\rm as}\ \ \  
m_{\sigma_8}^{\rm theor} = 0.88\ {\rm GeV},\ \ \ \ \ \ 
\label{eq:val}
\end{eqnarray}
where we have used the experimental values $m_\kappa$=0.91 GeV 
and $m_{a_0(980)}$=0.98 GeV. This value of 
$m_{\sigma_8}^{\rm theor}$ is quite close 
to $m_{\sigma_8}$=0.87 GeV, obtained
phenomenologically in Eq. (\ref{eq:mm}).
This fact supports our classification that ${\sigma}${\bf (600),} 
${\bf f_0}${\bf (980),}
${\kappa}${\bf (900) and }
${\bf a_0(980)}$ 
{\bf form a single scalar nonet.}\cite{ref2}  



\section{Chiral symmetry and properties of
the $\sigma$-nonet}


\begin{figure}[t]
 \epsfysize=7.0 cm
 \centerline{\epsffile{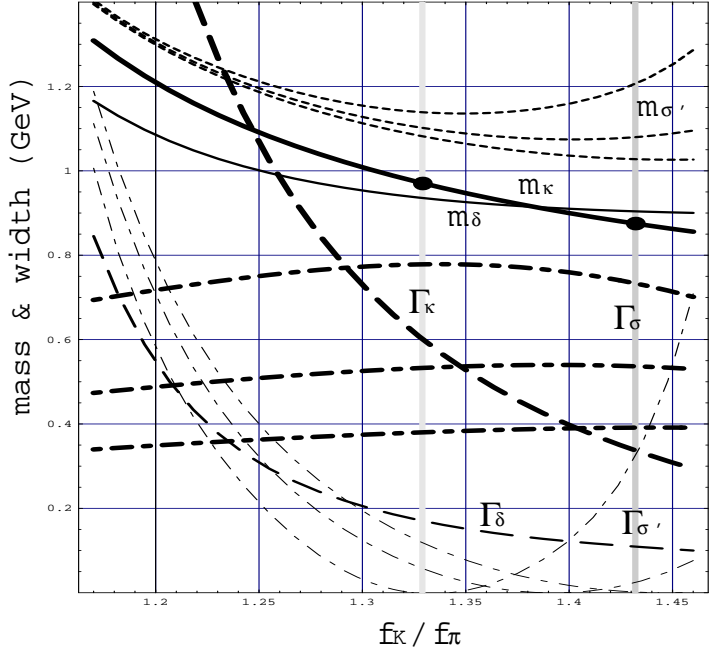}}
 \caption{The scalar meson masses and widths (GeV) 
versus $f_K/f_\pi$.
The upper, middle and lower lines of $M_\sigma '$, $\Gamma_\sigma$
and $\Gamma_{\sigma '}$ 
correspond, respectively, 
to the input 
values $M_\sigma =650,\ 585$ and 535 MeV.
}
\label{fig:tree}
\end{figure}

We assume that our scalar $\sigma$-nonet is a composite 
$q\bar q$-system as a chiral partner of the pseudoscalar 
$\pi$-nonet and that 
in the low energy region, where the 
structure of composite mesons can be neglected, 
they can $effectively$ be described by 
the linear $\sigma$ model (L$\sigma$M).
In the matrix 
notation $B\equiv s +i\phi$ ($s\equiv\lambda^i s^i/\sqrt{2}$ and 
$\phi\equiv\lambda^i\phi^i/\sqrt{2}$
denoting the scalar and pseudoscalar meson nonet, respectively),
the Lagrangian 
of $SU(3)$L$\sigma$M\cite{rf:gg,ref2,ref3,ref5} is
\begin{eqnarray}
{\cal L}^{{\rm L}\sigma{\rm M}} &=& 
\frac{1}{2}\langle \partial_\mu B\partial^\mu B
^\dagger 
\rangle 
-\frac{\mu^2}{2}\langle BB
^\dagger 
\rangle
-\frac{\lambda_1}{4}\langle BB^\dagger \rangle^2
-\frac{\lambda_2}{2}\langle (BB^\dagger )^2\rangle
\nonumber\\  
 &+& \kappa_d({\rm det}\ B+{\rm det}\ B^\dagger )
+\langle f s \rangle ,
\label{eq:lsm3}
\end{eqnarray}
where $\langle\ \ \rangle$ represents the trace.
Here $f$ is proportional to the current-quark mass matrix in QCD
and the form  
 $f={\rm diag}\{ f_n,f_n,f_s\} $ guarantees the PCAC.
In the process of spontaneous chiral symmetry breaking,
$s$ acquires the vacuum expectation value 
$s_0\equiv \Sigma ={\rm diag}\{ a,a,b\}$, and 
$s\phi\phi$-couplings appear. The 
pseudoscalar decay constants $f_\pi$ and $f_K$,
and their ratio are represented by 
\begin{eqnarray}
f_\pi &=& \sqrt{2}a,\ 
f_K=\frac{a+b}{\sqrt{2}};\ \ 
\frac{f_K}{f_\pi}
=\frac{a+b}{2a}. 
\label{eq:fp}
\end{eqnarray}
The six model parameters
contained in Eq. (\ref{eq:lsm3}) 
are determined by
the masses of $\pi ,\eta , \eta '$, $\sigma$ and $\kappa$ , 
and the decay constant $f_\pi$, and thus
we can predict the masses and widths
 of the scalar mesons. 
These are given in Table \ref{tab:tree}. 
The predicted properties 
are very sensitive to the value of 
$f_K/f_\pi$,\cite{rf:gg} 
as shown in Fig. \ref{fig:tree}.\cite{ref2}
The deviation of the value of $f_K/f_\pi$
from 1 represents the degree of $SU(3)$ breaking 
by $s_0$, as can be seen from Eq. (\ref{eq:fp}). 
We prefer the region in which  
this ratio satisfies $1.329<f_K/f_\pi<1.432$ (which is somewhat 
larger than the experimental value 1.22)
indicated by the  two vertical lines in the figure,
where the value $m_\kappa^{\rm theor}$ 
reproduces the experimental value 
within its uncertainty.
In this region, $\Gamma_\sigma$ and $\Gamma_\kappa$ 
are obtained with much larger values than those of 
$\Gamma_{\sigma '}$ and $\Gamma_\delta$. 
The reason that $\Gamma_{\kappa \rightarrow K\pi }$,
in spite of its comparatively smaller phase space,
becomes as large as $\Gamma_{\sigma \rightarrow\pi\pi }$
is due to the contribution 
to the coupling constant $g_{\kappa  K\pi}$  
from the determinant-type interaction in Eq. (\ref{eq:lsm3}).
The predicted widths of $\sigma$ and $\kappa$
are consistent with the experimental values. 
The predicted masses and widths of the other members,
$\delta (I=1)$ and $\sigma '(s\bar{s})$, are
close to those of
$a_0(980)$ and $f_0(980)$, respectively. 
\begin{table}
\begin{center}
\caption{The properties of the scalar meson nonet
predicted by $SU(3)$L$\sigma$M, compared with experiments. 
The underlined values of $m_\sigma$ and $m_\kappa$
along with $f_\pi$, $m_\pi$, 
$m_\eta$ and $m_{\eta '}$ are used as inputs.
The region of the value of $m_\kappa^{\rm exp}$ corresponds to 
the region in which the ratio of the  
decay constants satisfies 
$1.329<f_K^{{\rm L}\sigma{\rm M}}/f_\pi^{{\rm L}\sigma{\rm M}}<1.432$.
The properties of $\delta$ and $\sigma '$
become close to those of the observed resonances
$a_0(980)$ and $f_0(980)$, respectively,
 taken as the experimental candidates. 
The quantity $\Gamma_{\sigma '}^{\rm theor}$ is the partial width
$\Gamma_{\sigma '\rightarrow\pi\pi}^{\rm theor}$.
The value of $\Gamma_{\sigma '\rightarrow KK}^{\rm theor}$
is highly dependent on $m_{\sigma '}^{\rm theor}$,
since $m_{\sigma '}^{\rm theor}$ is close to 
$K\bar K$-threshold.
 }
\begin{tabular}{lcccc}
\hline
\hline
   & $m^{\rm theor}$/MeV & $m^{\rm exp}$/MeV
   & $\Gamma^{\rm theor}$/MeV & $\Gamma^{\rm exp}$/MeV\\
\hline
  $\sigma$ & $\underline{535\sim 650}$ & $\underline{535\sim 650}$
           & $400\sim 800$       & $385\pm 70$ \\
\hline
  $\kappa$ & $\underline{905\stackrel{+65}{\scriptstyle -30}} $
           & $\underline{905\stackrel{+65}{\scriptstyle -30}}$  
   & $300\sim 600$ & $545\stackrel{+235}{\scriptstyle -110} $\\
\hline
  $\delta =a_0(980)$ & $900\sim 930$ & $982.7\pm 2.0$  
           & $110\sim 170$ & $57\pm 11$ \\
\hline
  $\sigma '=f_0(980)$ & $1030\sim 1200$  & $993.2\pm 9.5$
             &  $0\sim 300$   & $67.9\pm 9.4$\\
\hline
\end{tabular}
\label{tab:tree}
\end{center}
\end{table}
The $\sigma '$ and $\delta$ states have large $K\bar K$-coupling
constants.\footnote{In the case $f_\pi =93$ MeV, $f_K/f_\pi =1.394$
and $m_\sigma =585$ MeV,
for $\sigma '$, 
${\cal L}_{\rm int}=g_{\sigma '\pi\pi}\sigma '{\pi}^2
+g_{\sigma 'KK}\sigma '(K^+K^-+K^0\bar K^0),
g_{\sigma '\pi\pi}=-0.02\ {\rm GeV},
\ {\rm and}\ g_{\sigma 'KK}=-4.97\ {\rm GeV}$.\\
For $\delta$, 
${\cal L}_{\rm int}=g_{\delta\pi\eta}
(\pi^-\delta^++\pi^+\delta^-+\pi^0\delta^0)\eta
+g_{\delta KK}(\delta^0\frac{K^+K^--K^0\bar K^0}{\sqrt 2}) 
+\delta^+K^0K^-+\delta^-\bar K^0K^+),\ 
g_{\delta\pi\eta}=-3.12\ {\rm GeV},
\ {\rm and}\ g_{\delta KK}=-3.19\ {\rm GeV}.$
}
(Especially the $\sigma '$ strongly couples
to the $K\bar K$ channel.)
This suggests that these states may also  
 be interpreted as 
$K\bar K$-molecule states.\cite{rf:Isg}\\
In L$\sigma$M, $\sigma $ and $\delta $
have almost the same quark content. Despite this fact,
$m_\delta$ becomes much larger than $m_\sigma$.
This phenomenon is explained by the properties of the  
instanton-induced $U_A(1)$-breaking determinant-type 
interaction.\\
%
Thus, it is promising to regard 
${\sigma}${\bf (600)}, ${\kappa}${\bf (900)}, 
{\bf a}${}_0${\bf (980)}, {\bf and} {\bf f}${}_0${\bf (980)}
{\bf as members of the scalar nonet,
forming with the members of} ${\pi}${\bf -nonet a linear 
representation of the SU(3) chiral symmetry}.\footnote{
This interesting assignment was suggested
and  insisted upon repeatedly by 
M. D. Scadron.\cite{rf:sca} 
}
\section{Repulsive background phase shift}
In our phase shift analyses of the $I=0$ $\pi\pi$ system 
the $\delta_{\rm BG}$
of hard core type  introduced phenomenologically 
played an essential role.
The strong cancellation between the positive $\delta_\sigma$ due to 
$\sigma$-resonance and the negative $\delta_{BG}$ leads to the 
existence of $\sigma$ meson.
In the $I=2$ $\pi\pi$ system, where there is
no known / expected resonance,
the phase shift of repulsive core type appears directly.\cite{ref1}\\
The 
origin of this $\delta_{\rm BG}$ seems to  have a close connection to 
 the $\lambda\phi^4$-interaction in L$\sigma$M\cite{ref2}:
It represents a contact zero-range interaction
and is strongly repulsive both in the $I=0$ and 2 systems,
and has plausible properties
as the origin of repulsive core.\\
The $\pi\pi$-scattering $A(s,t,u)$-amplitude  by $SU(2)$ L$\sigma$M 
consists of the amplitude due to $\sigma$-production, which is
strongly attractive, and the amplitude due to contact $\lambda\phi^4$
interaction, which is strongly repulsive. 
These two terms cancel each other
in $O(p^0)$ level, and 
the $O(p^2)$ Tomozawa-Weinberg amplitude
and the $O(p^4)$ (and higher order) correction 
term are left.
As a result the derivative coupling property
of $\pi$-meson as a Nambu-Goldstone boson is
preserved.
In this sense the $\lambda\phi^4$-interaction can be
called a ``compensating" interaction for
$\sigma$-effect.\\
Thus the strong cancellation
between the positive $\delta_\sigma$  
and the negative $\delta_{\rm BG}$ in our analysis 
 is reducible to this cancellation in $A(s,t,u)$ amplitude 
guaranteed by chiral symmetry.\footnote{
Similar cancellation mechanism also acts in $K\pi$ 
scattering amplitude, leading to existence of $\kappa$ resonance,
and the origin of $\delta_{BG}$ in  $I=1/2$ channel
is explained by the compensating repulsive interaction.
}\\
We can directly estimate the magnitude of $\delta_{BG}$ in the framework
of L$\sigma$M by using N/D method. The behavior of $\delta_{BG}$
is almost consistent with the theoretical prediction by the L$\sigma$M
including $\rho$ meson contribution.\cite{ref4}
\section{Concluding remark}
In this talk first I have pointed out that 
the scalars below 1 GeV, $\sigma (600)$,
$\kappa (900)$, $a_0(980)$ and $f_0(980)$ 
 are possibly to form a single scalar nonet\cite{ref2,ref3}: 
The octet members of this nonet satisfies the 
Gell-Mann Okubo mass formula.
The masses and widths 
of the scalar $\sigma$-nonet are also shown to 
satisfy the relation predicted by 
$SU(3)$ L$\sigma$M. This fact suggests the linear representation of 
$SU(3)$ chiral symmetry is realized in nature.\\
Secondly I have reported on the recent results concerning the 
background phase shift  $\delta_{BG}$:
In our phase shift analysis there occurs a strong cancellation between
$\delta_\sigma$ due to the $\sigma$ resonance 
and $\delta_{BG}$, which is guaranteed by 
chiral symmetry.
A reason of overlooking the $\sigma$ in conventional phase shift analysis
seems to be due to overlooking of this cancellation mechanism.
The  phenomenological behavior of $\delta_{BG}$ is   
quantitatively describable in the framework of L$\sigma$M including
$\rho$ meson contribution.\\
Finally I add some comments on the related works.
The authors in Ref.\cite{ref6} have insisted that 
the above-mentioned  scalar $\sigma$-nonet is
to be  $qq\bar q\bar q$ states, 
being based on the results of their analysis; 
the $\gamma\gamma$ widths of $a_0$ and $f_0$,
assuming them to be $^3P_0$ $q\bar q$ states,
are much smaller than the theoretical predictions, obtained by using
the $q\bar q$ model, from 
the corresponding widths of  $a_2$ and $f_2$, 
the established $^3P_2$ $q\bar q$ states, respectively, 
However, 
the $\sigma$-nonet is to be discriminated from the conventional
$^3P_0$ $q\bar q$ state, and is to be regarded as 
a relativistic (positive parity) ``$S$-wave'' $q\bar q$ state 
from the viewpoint of linear representation of 
chiral symmetry.\cite{refS}
Actually, the L$\sigma$M prediction for the $2\gamma$ decay rates of
the $a_0$, $f_0$ mesons
is shown to agree
with experimental data.\cite{ref7} \\
\end{document}